\documentclass[12pt]{article}
\usepackage{epsfig}
\usepackage{wrapfig}
\usepackage{subfigure}
\usepackage{amsmath}
\usepackage{hhline}
\usepackage{amssymb}
\usepackage{times}
\usepackage{cite}

\usepackage[]{lineno}

\newlength{\dinwidth}
\newlength{\dinmargin}
\begin{document}  
\newcommand{\pom}{{I\!\!P}}
\newcommand{\reg}{{I\!\!R}}
\newcommand{\slowpi}{\pi_{\mathit{slow}}}
\newcommand{\fiidiii}{F_2^{D(3)}}
\newcommand{\fiidiiiarg}{\fiidiii\,(\beta,\,Q^2,\,x)}
\newcommand{\n}{1.19\pm 0.06 (stat.) \pm0.07 (syst.)}
\newcommand{\nz}{1.30\pm 0.08 (stat.)^{+0.08}_{-0.14} (syst.)}
\newcommand{\fiidiiiful}{F_2^{D(4)}\,(\beta,\,Q^2,\,x,\,t)}
\newcommand{\fiipom}{\tilde F_2^D}
\newcommand{\ALPHA}{1.10\pm0.03 (stat.) \pm0.04 (syst.)}
\newcommand{\ALPHAZ}{1.15\pm0.04 (stat.)^{+0.04}_{-0.07} (syst.)}
\newcommand{\fiipomarg}{\fiipom\,(\beta,\,Q^2)}
\newcommand{\pomflux}{f_{\pom / p}}
\newcommand{\nxpom}{1.19\pm 0.06 (stat.) \pm0.07 (syst.)}
\newcommand {\gapprox}
   {\raisebox{-0.7ex}{$\stackrel {\textstyle>}{\sim}$}}
\newcommand {\lapprox}
   {\raisebox{-0.7ex}{$\stackrel {\textstyle<}{\sim}$}}
\def\gsim{\,\lower.25ex\hbox{$\scriptstyle\sim$}\kern-1.30ex%
\raise 0.55ex\hbox{$\scriptstyle >$}\,}
\def\lsim{\,\lower.25ex\hbox{$\scriptstyle\sim$}\kern-1.30ex%
\raise 0.55ex\hbox{$\scriptstyle <$}\,}
\newcommand{\pomfluxarg}{f_{\pom / p}\,(x_\pom)}
\newcommand{\dsf}{\mbox{$F_2^{D(3)}$}}
\newcommand{\dsfva}{\mbox{$F_2^{D(3)}(\beta,Q^2,x_{I\!\!P})$}}
\newcommand{\dsfvb}{\mbox{$F_2^{D(3)}(\beta,Q^2,x)$}}
\newcommand{\dsfpom}{$F_2^{I\!\!P}$}
\newcommand{\gap}{\stackrel{>}{\sim}}
\newcommand{\lap}{\stackrel{<}{\sim}}
\newcommand{\fem}{$F_2^{em}$}
\newcommand{\tsnmp}{$\tilde{\sigma}_{NC}(e^{\mp})$}
\newcommand{\tsnm}{$\tilde{\sigma}_{NC}(e^-)$}
\newcommand{\tsnp}{$\tilde{\sigma}_{NC}(e^+)$}
\newcommand{\st}{$\star$}
\newcommand{\sst}{$\star \star$}
\newcommand{\ssst}{$\star \star \star$}
\newcommand{\sssst}{$\star \star \star \star$}
\newcommand{\tw}{\theta_W}
\newcommand{\sw}{\sin{\theta_W}}
\newcommand{\cw}{\cos{\theta_W}}
\newcommand{\sww}{\sin^2{\theta_W}}
\newcommand{\cww}{\cos^2{\theta_W}}
\newcommand{\trm}{m_{\perp}}
\newcommand{\trp}{p_{\perp}}
\newcommand{\trmm}{m_{\perp}^2}
\newcommand{\trpp}{p_{\perp}^2}
\newcommand{\alp}{\alpha_s}
\newcommand{\etamax}{\eta_{\rm max}}

\newcommand{\alps}{\alpha_s}
\newcommand{\sqrts}{$\sqrt{s}$}
\newcommand{\LO}{$O(\alpha_s^0)$}
\newcommand{\Oa}{$O(\alpha_s)$}
\newcommand{\Oaa}{$O(\alpha_s^2)$}
\newcommand{\PT}{p_{\perp}}
\newcommand{\JPSI}{J/\psi}
\newcommand{\sh}{\hat{s}}
\newcommand{\uh}{\hat{u}}
\newcommand{\MP}{m_{J/\psi}}
\newcommand{\PO}{I\!\!P}
\newcommand{\xbj}{x}
\newcommand{\xpom}{x_{\PO}}
\newcommand{\ttbs}{\char'134}
\newcommand{\xpomlo}{3\times10^{-4}}  
\newcommand{\xpomup}{0.05}  
\newcommand{\dgr}{^\circ}
\newcommand{\pbarnt}{\,\mbox{{\rm pb$^{-1}$}}}
\newcommand{\gev}{\,\mbox{GeV}}
\newcommand{\WBoson}{\mbox{$W$}}
\newcommand{\fbarn}{\,\mbox{{\rm fb}}}
\newcommand{\fbarnt}{\,\mbox{{\rm fb$^{-1}$}}}
\newcommand{\dsdx}[1]{$d\sigma\!/\!d #1\,$}
\newcommand{\eV}{\mbox{e\hspace{-0.08em}V}}
%
%
\newcommand{\qsq}{\ensuremath{Q^2} }
\newcommand{\gevsq}{\ensuremath{\mathrm{GeV}^2} }
\newcommand{\et}{\ensuremath{E_t^*} }
\newcommand{\rap}{\ensuremath{\eta^*} }
\newcommand{\gp}{\ensuremath{\gamma^*}p }
\newcommand{\dsiget}{\ensuremath{{\rm d}\sigma_{ep}/{\rm d}E_t^*} }
\newcommand{\dsigrap}{\ensuremath{{\rm d}\sigma_{ep}/{\rm d}\eta^*} }

\newcommand{\dstar}{\ensuremath{D^*}}
\newcommand{\dstarp}{\ensuremath{D^{*+}}}
\newcommand{\dstarm}{\ensuremath{D^{*-}}}
\newcommand{\dstarpm}{\ensuremath{D^{*\pm}}}
\newcommand{\zDs}{\ensuremath{z(\dstar )}}
\newcommand{\Wgp}{\ensuremath{W_{\gamma p}}}
\newcommand{\ptds}{\ensuremath{p_t(\dstar )}}
\newcommand{\etads}{\ensuremath{\eta(\dstar )}}
\newcommand{\ptj}{\ensuremath{p_t(\mbox{jet})}}
\newcommand{\ptjn}[1]{\ensuremath{p_t(\mbox{jet$_{#1}$})}}
\newcommand{\etaj}{\ensuremath{\eta(\mbox{jet})}}
\newcommand{\detadsj}{\ensuremath{\eta(\dstar )\, \mbox{-}\, \etaj}}

\def\Journal#1#2#3#4{{#1} {\bf #2} (#3) #4}
\def\NCA{\em Nuovo Cimento}
\def\NIM{\em Nucl. Instrum. Methods}
\def\NIMA{{\em Nucl. Instrum. Methods} {\bf A}}
\def\NPB{{\em Nucl. Phys.}   {\bf B}}
\def\PLB{{\em Phys. Lett.}   {\bf B}}
\def\PRL{\em Phys. Rev. Lett.}
\def\PRD{{\em Phys. Rev.}    {\bf D}}
\def\ZPC{{\em Z. Phys.}      {\bf C}}
\def\EJC{{\em Eur. Phys. J.} {\bf C}}
\def\CPC{\em Comp. Phys. Commun.}

\begin{titlepage}

\noindent

\noindent
\noindent

\vspace{2cm}
\begin{center}
\begin{Large}

{\bf Statistical Methods for Estimating the non-random Content 
of Financial Markets 
}
\end{Large}

\vspace{2cm}

Laurent Schoeffel \\~\\
CEA Saclay, Irfu/SPP, 91191 Gif/Yvette Cedex, \\
France
\end{center}

\vspace{2cm}

\begin{abstract}

For the pedestrian observer, financial markets look completely random with erratic and uncontrollable
behavior. To a large extend,  this is correct.
At first approximation the difference 
between real price changes and the random walk model
is too small to be detected using
traditional time series analysis.
However, we show in the following that this difference between real financial time series and
random walks, as small as it is, is detectable using modern statistical multivariate analysis,
with several triggers encoded in trading systems.
This kind of analysis are based on methods widely used in nuclear physics, with large samples
of data and advanced statistical inference.
Considering the movements of the 
Euro future
contract at high frequency,
we show that a part of the non-random content of this series can be inferred,
namely the trend-following content depending on volatility ranges.

\end{abstract}

\vspace{1.5cm}

\begin{center}
\end{center}

\end{titlepage}

%
%
%
%

\section{Introduction}

The random walk model of price changes in financial time series has been 
so durable because it is nearly correct. At first approximation the difference 
between real price changes and the random walk model
is too small to be detected using
traditional time series analysis \cite{MS,r2,r3}.
More precisely, when looking at large samples of data, some features appear
that break the  random walk approximation. For example, the 
statistics of price distribution at small time scales is not Gaussian but
governed by non-extensive statistics \cite{ts1,ts2}. We can also detect large range correlation
in the absolute returns, which mean that persistent behaviors exist that are
not embedded in the random walk model \cite{MS,r2,r3}, which can be seen as a consequence of the
non-extensive statistics \cite{ts3}. 

However, any deviations from the Gaussian limit 
can not be detected at a local level, 
when we observe the market in a short window of time.  Our first statement is still valid. At first order of observation, prices
in financial markets
behave randomly and it remains impossible to predict whether the next price
movement will be up or down.  

We show in the following that this difference between real markets and
random walks, as small as it is, is detectable using modern statistical analysis
with hypothesis testing, even when we observe the market locally. In particular, it is detectable once we wan build a
trading systems on the basis of multivariate analysis and hypothesis testing \cite{stat1,stat2}.
Indeed,
tools of statistical physics have been proven to be efficient in many areas,
like extracting the average properties of a macroscopic 
system from its microscopic dynamics, even if approximately known.
The same  holds for financial systems.
Even though it is difficult or almost impossible to write down 
the microscopic equation of motion that drives prices at each instant,
it is possible to extract a relevant statistical information, that makes sense to 
take decisions at a local level.
That's what we exemplify in this article on the behavior with time of
the Euro future
contract (EC) at high frequency.
We show that we can infer the non-random content of the EC erratic behavior
using a multivariate analysis embedded in a trading algorithm.

\section{Data sets and Data treatment}

We use five minutes sampling of the EC time series, from January 2000 till August 2011,
which makes 839k quotes that we use to build the trading system. We 
conserve only the close of each quote. This large sample of data points is necessary
to infer statistical properties with a high confidence level, as shown in the following. 
Also, in the context
of this analysis, the fine tuning of the time series with a five minutes resolution
is useful to focus on possible intermittent behavior of the series at small scales (five minutes),
that could disappear at larger scales.

A typical quote of the EC is like $1.3802$. The unit of the last digit is what we call a basis point. For example, we consider that
a price movement from $1.3802$ to $1.3805$ corresponds to a price change
of $3$ basis points. More precisely, if we buy the contract at time T1 (on the quote Q1)
at $1.3802$ and sell this contract at time T2$>$T1 (on the quote Q2) at $1.3805$, then this
trade corresponds to a gain of $3$ basis points (without fees). To keep the procedure as
close to reality as possible we consider fees of two times the slippage, which means that
this trade is counted in our approach as a trade of $1$ basis point (net of fees).

A fundamental issue in the analysis is to break the data samples in three parts, that we call 
in-sample, out-sample and live-sample. The decomposition is done as follows:
\begin{itemize}
\item[(i)] 2000-2007: in-sample
\item[(ii)] 2008-2009: out-sample
\item[(iii)] 2010-2011: live-sample
\end{itemize}
What is the interest of this decomposition of the data series? The idea is that we intend to build a 
trading system on this series. This means that we intent to design an algorithm 
that will take decisions like buy or sell 1 EC contract at a given quote.
This decision at a given quote will be based on multivariate analysis, as mentioned in the beginning.
In order to process  this way, we need a data sample on which the algorithm is built
and all parameters of the algorithm are fitted.
Therefore, this data sample needs to be large in order to be relevant statistically. 
This sample is called in-sample (i) and 
is defined as the period 2000-2007.

The second sample, called out-sample (ii), is used as a  validation stage.
All algorithms built on (i) are expected obviously to give satisfactory results on (i). However, as parameters of
the model are fitted on the sample (i), there is no guarantee that the model could behave properly on another data sample.
If it does so, this means that the algorithm is not a pure artifact and contains a part of the real dynamics of the
market. 
This is the purpose of the sample (ii), defined as the period 2008-2009.
If the trading system built on (i) fails on (ii), it is rejected and another algorithm is designed.
Note that we have other intermediate validation stage to make the full process more robust: we come back on this point
later  in the article.
Also, note that there is no guarantee at this level that what we describe in this paragraph is possible.

Finally, once we have obtained an algorithm that works on (ii) and satisfies our robustness tests,
if any, we test it on what we call the live-sample (iii), defined as the period 2010-2011. Our building process is made to guarantee at this step the good functioning of the trading system and that's what we show in the following.

Note that if we can drive the analysis to this last step and if it works, it is a clear proof of our claim of the 
previous part on a specific example (EC):
the difference between real markets and
random walks, as small as it is, is detectable using modern statistical analysis
in multivariate analysis. The multivariate approach refers to 
the number of parameters introduced in the definition of triggers for trades decisions 
along the EC series.

\section{Strategy Reconstruction}

\begin{figure}[htbp]
  \begin{center}
    \includegraphics[width=1.\textwidth]{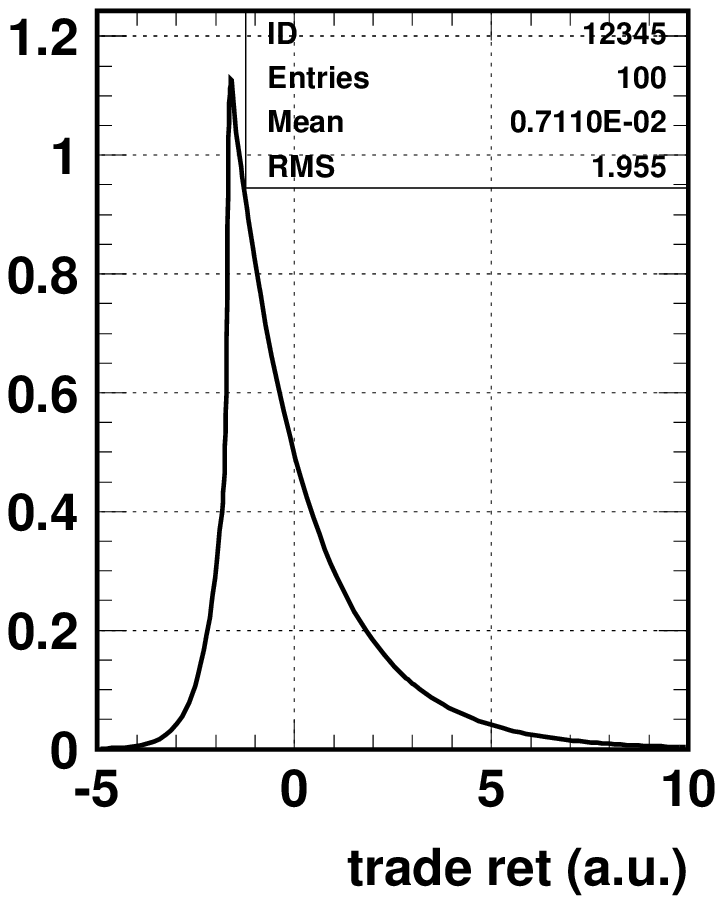}
  \end{center}
  \caption{Trade return distribution for a theoretical trend-following system (see text).}
\label{theo}
\end{figure}

The basic elements of our approach are much simple. 
The gross feature of the strategy we explore is a trend-following mechanism \cite{bo}.
The trade distribution for this simple theoretical system is given in Fig. \ref{theo}.
Obviously, there is no possibility with such a simple algorithm to reconstruct a profitable strategy
on ten years of high frequency data.
Then, we use  the idea of  trend-following, in building a
more complex structure based on additional sub-triggers on volatilities.  Therefore, 
if we can show that this strategy leads to profitable results (net of fees), it will be a proof of
the validity of the trend-following hypothesis on the market, taking into account multivariate tests to
activate the trend follower.
More specific details of the model will be provided in a further publication.
Let us note that the use of moving averages is a powerful experimental method to access to the non-trivial 
statistical texture of a time series. If we consider 2 standard moving averages of lengths $T_1$
and $T_2>T_1$, with $\Delta T = (T_2-T_1)/T_1$, then the density of crossing points of the 2 averages
is given by:
\begin{equation}
\rho \sim \frac{1}{T_2} [\Delta T (1-\Delta T)]^{H-1}
\end{equation}
where $H$ is the Hurst exponent, that characterizes the persistence or anti-persistent of the
data series \cite{hurst}.

\begin{figure}[htbp]
  \begin{center}
    \includegraphics[width=0.8\textwidth]{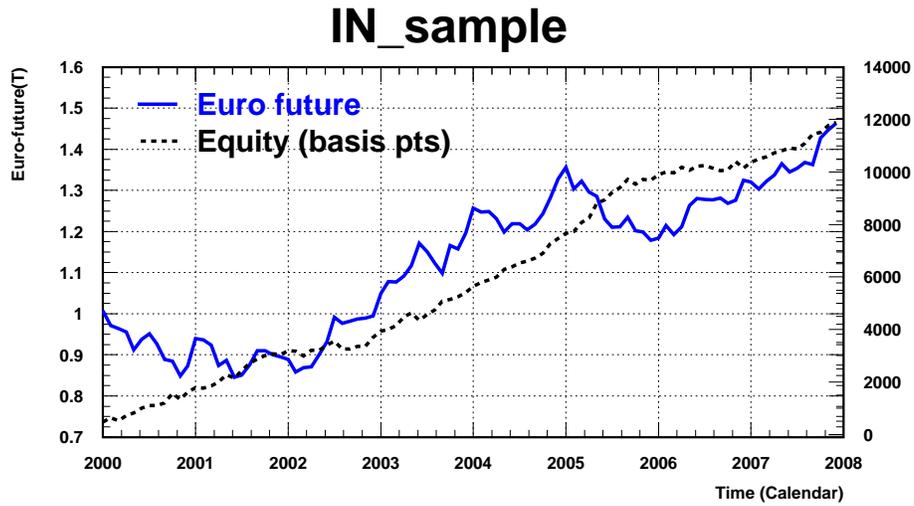}
    \includegraphics[width=0.8\textwidth]{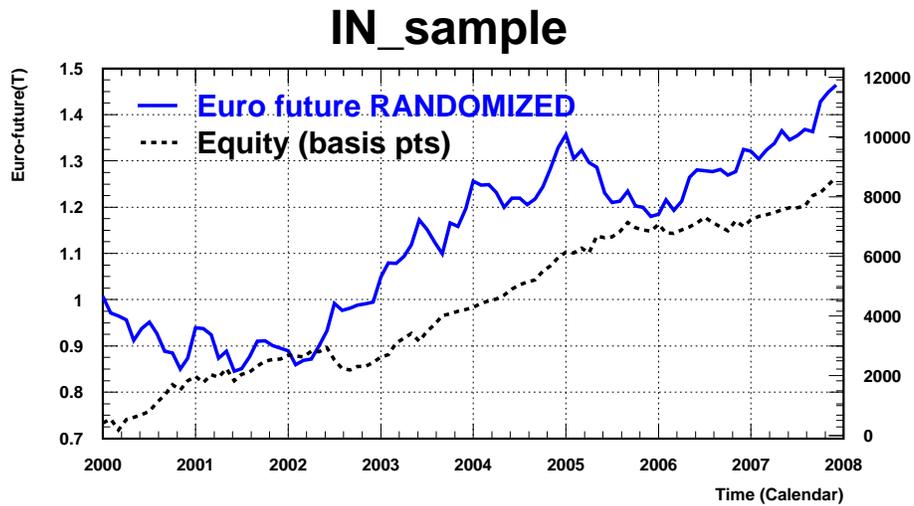}
  \end{center}
  \caption{Behavior with time of the EC contract on the in-sample
part (i) of the data series (full line) as well as the cumulative equity (dashed line)
of the designed trading system (expressed in basis points on the right vertical axis, net of fees).
The bottom plot corresponds also to the running of the trading system,
but this time on the randomized in-sample. See text for details.}
\label{in}
\end{figure}

We have $8$ parameters optimized on the in-sample (i). 
The optimization is performed in order to achieve the best Sharpe ratio.
Results are shown in Fig. \ref{in}.
We present the behavior with time of the EC contract itself as well as the cumulative equity
of the designed trading system (expressed in basis points, net of fees).
We observe the nice behavior of the equity, increasing with time, which shows that
the strategy is profitable and coherent with respect to different market regimes.
The bottom plot in Fig. \ref{in} corresponds also to the running of the trading system,
but this time on the randomized in-sample. Exactly, we have added to each quote value of
the data series (i) a random number that ranges between $-10$ and $+10$ times the slippage of the EC contract.
And we run the trading system on this series, which leads to the bottom plot of
Fig. \ref{in}.
This randomization is necessary as we do not want the trading system to be dependent on the
point-to-point correlation and also the model must be flexible to absorb distortion of the
data series. This is what we observe in Fig. \ref{in} (bottom): the system is robust against
randomization of the data series. Note that all systems designed and that have
failed at this stage have been rejected.

\begin{figure}[htbp]
  \begin{center}
    \includegraphics[width=0.8\textwidth]{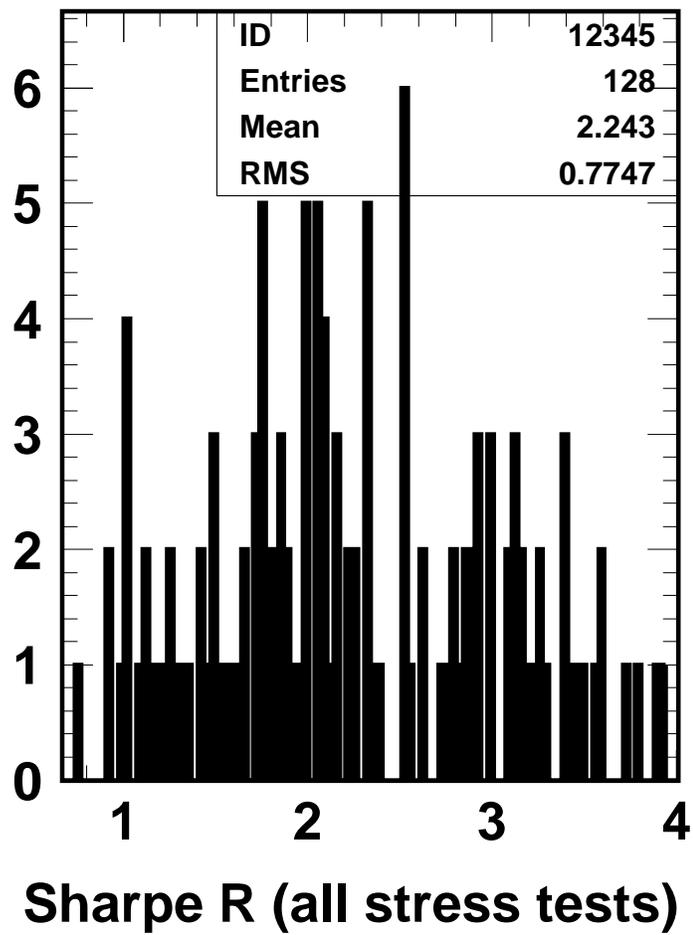}
  \end{center}
  \caption{
Spread of Sharpe ratios corresponding to all stress tests considered
in order to ensure the robustness of the trading system
(see text).
}
\label{stress}
\end{figure}

Before considering the out-sample stage, we have an intermediate essential step of 
validation of the trading system.
To ensure that the system is robust, we need more that the randomization of the data series.
We need to distort the strategy itself in many ways: for example, force an exit of 
given trade at a given time, do not execute randomly some trades, delay the execution of 
orders by several quotes, execute an order but at a wrong price, with a prejudice for the
trading system, multiply the fees (slippage) by a factor 2,3 or 4 etc. 
Thus, we have a list of  stress tests and for each case, 
the trading system is run and a result is obtained.
All this must be done on the original data series (i) and on its
randomized version. In all cases, we must observe that the system is stable and robust.
This is shown in Fig. \ref{stress}, where we present the Sharpe ratio for all stress tests
considered. We do not provide the equity in each configuration. We summarize each case by one
entry in Fig. \ref{stress}, as a value of the Sharpe ratio for the case under study.
The idea is that the robustness is ensured if we do not observe pathological values
in the Sharpe ratios, even for the more extreme stress tests. This is what we observe
 in Fig. \ref{stress}, with an average value of $2.2$ and a RMS of $0.8$. In all
configurations, the model stays reasonable. By this method, we have also shown that
the trading system does not depend on the fine tuning of any of the fitted parameters. Otherwise, 
a few stress tests would have failed deeply. On the contrary, our strategy depends weakly on any
of its inputs, which gives a lot of flexibility on all variables of the system with always
a profitable result obtained.

\begin{figure}[htbp]
  \begin{center}
    \includegraphics[width=0.8\textwidth]{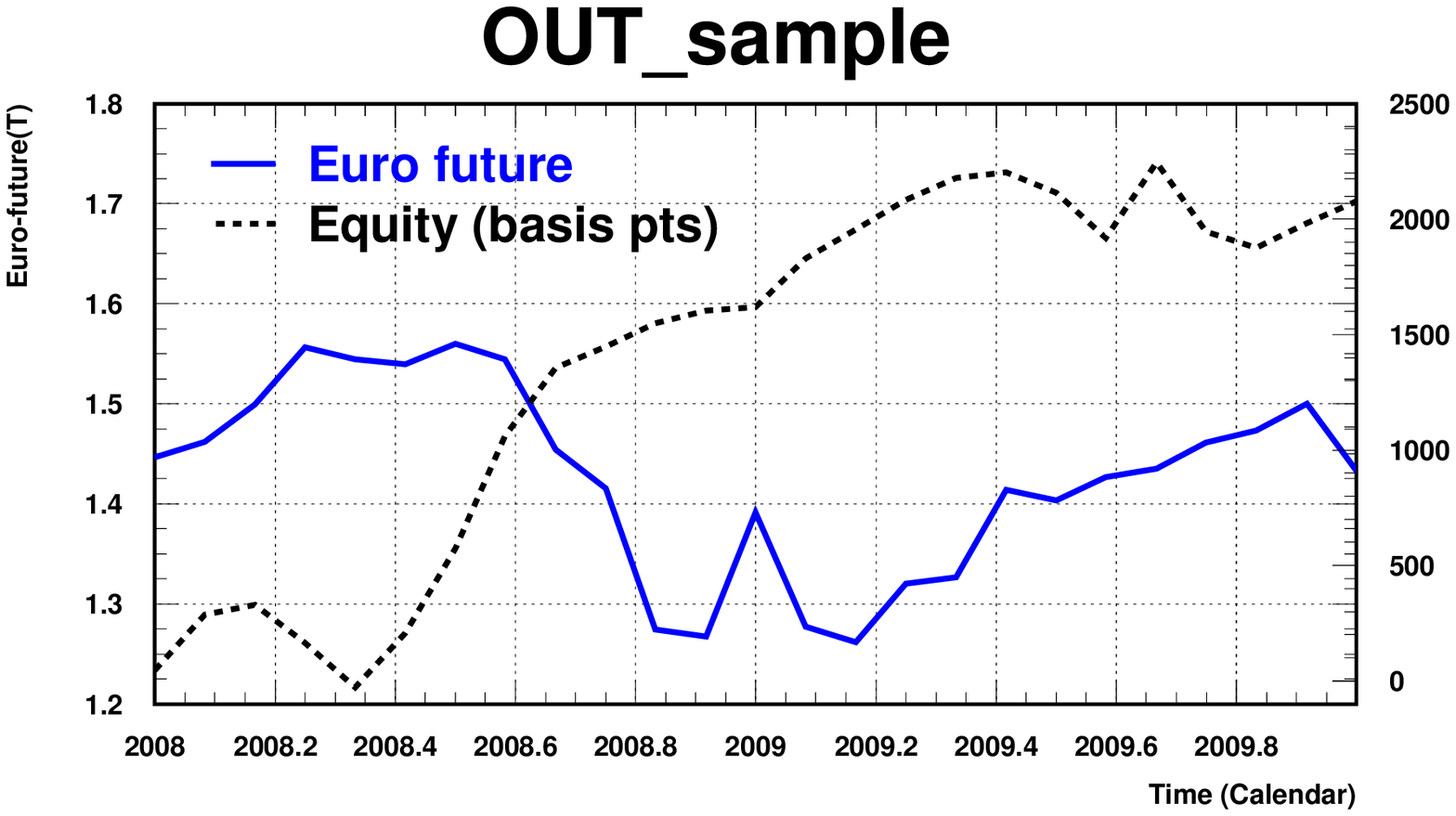}
  \end{center}
  \caption{Behavior with time of the EC contract on the out-sample
part (ii) of the data series (full line) as well as the cumulative equity (dashed line) 
of the designed trading system (expressed in basis points on the right vertical axis, net of fees).}
\label{out}
\end{figure}

At this step, it is not unreasonable to claim that we have designed a robust algorithm.
However, a new validation stage is determinant using the out-sample (ii).
This is a decisive test as we are running on new data, that the system does not know, in
the sense that parameters have been fitted on another set of data. In principle parameters
are robust as we have already explored many configurations for the data series and
the system. However, the out-sample test will kill all systems that still have some
elements of over-fit in their construction.
Indeed, such systems fail to give good results when running on the sample (ii) and are rejected. 
This is what happens for most of the systems that can be designed if the input ideas are not
carrying decisive features of the inside dynamics of the time series.
That's why it is not an easy task and many attempts are needed before converging towards
acceptable solutions.
In Fig. \ref{out} we show the result for the trading system described above.
We observe a correct behavior of the equity, which qualifies definitely this system.
 We interpret this as a clear 
evidence that the time series of the EC exhibits features of trend-following, under
certain ranges of volatilities, as encoded in our trading system.

\begin{figure}[htbp]
  \begin{center}
    \includegraphics[width=0.8\textwidth]{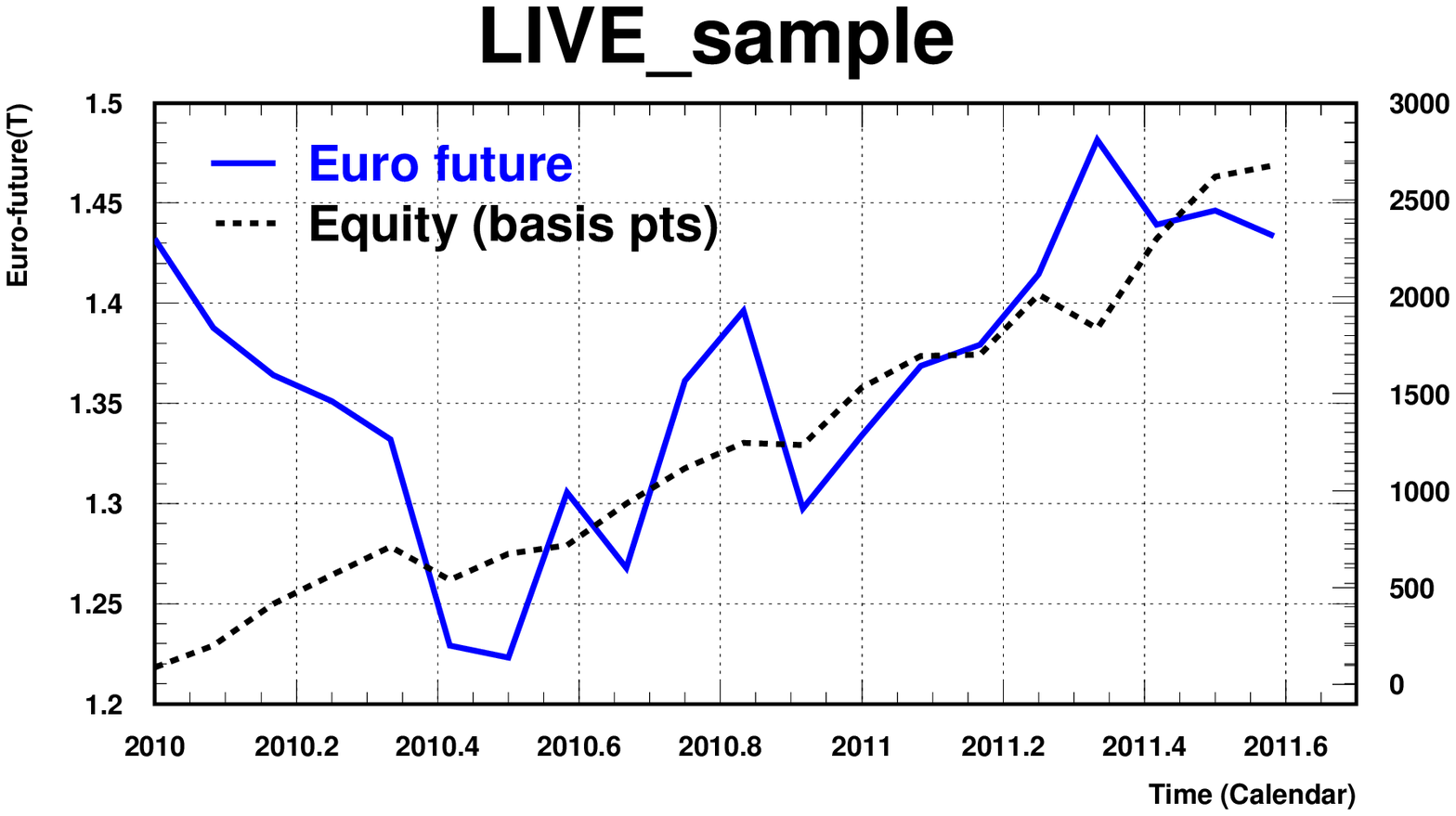}
  \end{center}
  \caption{Behavior with time of the EC contract on the live-sample
part (iii) of the data series (full line) as well as the cumulative equity (dashed line) 
of the designed trading system (expressed in basis points on the right vertical axis, net of fees).}
\label{live}
\end{figure}

Finally, in Fig. \ref{live}, we check the result on the live-sample period (iii),
in 2010-2011. Here, we do not expect any failure, otherwise the full process described
above must be rejected. Effectively, we observe a nice behavior of the cumulative
equity (net of fees), much compatible with what has been designed on the in-sample. 
This confirms our statement above on the dynamical content of the trading algorithm
we have presented here.

\begin{figure}[htbp]
  \begin{center}
    \includegraphics[width=0.45\textwidth]{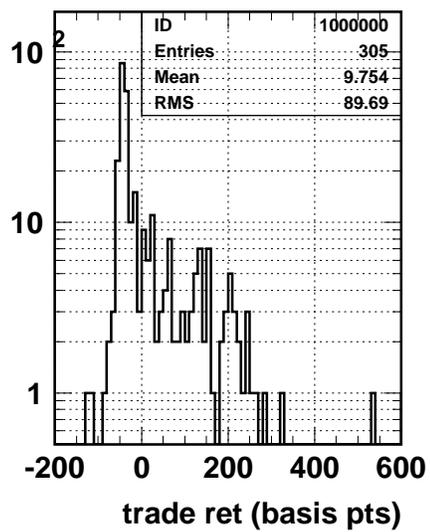}
    \includegraphics[width=0.45\textwidth]{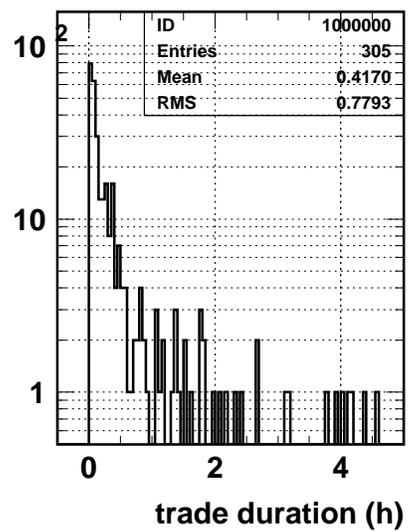}
  \end{center}
  \caption{Left: Trade return distribution for the live-sample (iii). Right: 
Trade duration for the live-sample (iii).}
\label{dist}
\end{figure}

In order to illustrate very simply the gross feature of the model, we present two distributions
in Fig. \ref{dist}. We show the trade return spectrum (Fig. \ref{dist}-left), in which we
recognize a typical trend-following system, reminiscent from the standard behavior
plotted in Fig. \ref{theo}. We observe also in Fig. \ref{dist} (right) that the system is
effectively working at high frequency with an average duration of trades of $25$ minutes.

\section{Conclusion}

Tools of statistical physics \cite{stat1,stat2} have been proven to be efficient in many scientific areas.
In a similar way for financial time series,
knowing that the difference between real markets and
random walks
is very small, a modern statistical multivariate analysis can 
help to extract this difference.
This is what is encoded in trading systems. We have shown how
to achieve the construction of such a system on the Euro future contract at high frequency.
A typical element of the dynamics of this system is then accessible,
namely the trend-following idea involved in a more complex structure
depending on volatilities.

An immediate question can be raised concerning the rationale behind this content.
Our observation is universal in the sense that the same algorithm is running on more that 10 years
of data, where the monetary policy has changed several times.
Then, this is not attached to a particular regime of interest rates.
There are certainly herding behaviors at the origin of the values of parameters encoded in
our system. These herding phases may appear with strengths governed by certain fear levels,
corresponding to  volatility ranges. Also, in some circumstances, nothing special can be said.
Finally, a global rationale explanation of a given trading system 
is very complex and probably not unique. This is beyond the scope of this article.


\newpage

\end{document}